\documentstyle[pra,aps,eqsecnum,epsf]{revtex}

\begin{document}
   \draft

\def\bx{\mbox{\boldmath $x$}}
\def\by{\mbox{\boldmath $y$}}
\def\bz{\mbox{\boldmath $z$}}
\def\ba{\mbox{\boldmath $a$}}
\def\bb{\mbox{\boldmath $b$}}
\def\bc{\mbox{\boldmath $c$}}
\def\bd{\mbox{\boldmath $d$}}
\def\bu{\mbox{\boldmath $u$}}
\def\bv{\mbox{\boldmath $v$}}
\def\bm{\mbox{\boldmath $m$}}
\def\bn{\mbox{\boldmath $n$}}
\def\br{\mbox{\boldmath $\rho$}}
\def\bs{\mbox{\boldmath $\sigma$}}
\def\bbm{\mbox{\boldmath $\mu$}}
\def\bH{\mbox{\boldmath $H$}}
\def\bE{\mbox{\boldmath $E$}}
\def\bxi{\mbox{\boldmath $\xi$}}

\def\hx{\widehat{\mbox{\boldmath $x$}}}
\def\hy{\widehat{\mbox{\boldmath $y$}}}
\def\hz{\widehat{\mbox{\boldmath $z$}}}
\def\hm{\widehat{\mbox{\boldmath $m$}}}
\def\hn{\widehat{\mbox{\boldmath $n$}}}
\def\hk{\widehat{\mbox{\boldmath $k$}}}

\title{Hamilton's Turns for the Lorentz Group} 
\author{R.Simon\thanks{email: simon@imsc.ernet.in}}
\address{ The Institute of 
Mathematical Sciences, C. I. T. Campus, Chennai 600 113, India} 
\author{S. Chaturvedi\thanks{e-mail: scsp@uohyd.ernet.in}}
\address{School of Physics, University of Hyderabad, Hyderabad 500 046, India}
\author{V. Srinivasan
\thanks {e-mail: vsspster@gmail.com}}
\address{Department of Theoretical Physics, University of Madras, Guindy Campus, 
Chennai 600 025}
\author{N. Mukunda\thanks{email: nmukunda@cts.iisc.ernet.in}}
\address{Centre for High Energy Physics, 
Indian Institute of Science, Bangalore 560 012, India}

\date{\today}

\maketitle

\begin{abstract}
Hamilton in the course of his studies on quaternions came up with an elegant 
geometric picture for the group $SU(2)$. In this picture  the group elements  are 
represented by ``turns'', which are equivalence classes of directed great circle 
arcs on the unit sphere $S^2$, in such a manner that  the rule for composition of 
group elements  
takes the form of the familiar parallelogram law for the Euclidean translation 
group. It is only recently that this construction has been generalized  to the 
 simplest noncompact 
group $SU(1,1) = Sp(2, R) =  SL(2,R)$, the double cover of $SO(2,1)$. The present 
work develops a theory of turns for $SL(2,C)$, the double and universal cover of 
$SO(3,1)$ and $SO(3,C)$, rendering a geometric representation in the 
spirit of 
Hamilton available for all low dimensional semisimple Lie groups of interest in 
physics. The geometric construction is illustrated through application to polar 
decomposition,  and to the composition of Lorentz boosts and the resulting 
Wigner or Thomas rotation.

\end{abstract}

\pacs{PACS: 02.20.-a}

\section{Introduction}
The group $SU(2)$ plays an important role in various branches of
physics. One is generally familiar with two ways of
parametrizing the elements of $SU(2)$ -- the Euler angle
parametrization and the Cayley-Klein parametrization. In neither
of these parametrizations is the expression for the group 
composition law particularly illuminating: there is no simple
way of remembering  or visualizing the composition law.  Hamilton\cite{hamilton},
 in the course of his studies on quaternions, developed an interesting   
 geometric picture for  representing the elements of 
$SU(2)$ wherein  the group composition law acquires the structure of the
familiar head to tail {\em parallelogram rule} of vector addition.
This work has come to be known as {\it Hamilton's theory of
turns}, and an excellent review can be found in the monograph of 
Biedenharn and Louck\cite{biedenharn}. Interestingly, this elegant geometric picture 
does not 
seem to be as well known as it deserves, and it is only recently that Hamilton's 
construction was
  generalized to 
the simplest noncompact semisimple Lie  group $SU(1,1)$\cite{prl,jmp}. 

The ideas underlying Hamilton's construction
can be best and most easily understood through analogy with the much simpler
case of the Abelian group of translations in a three dimensional
Euclidean space. Each element of this group can be thought of as
a unique point in a three dimensional space or, equivalently, as a
vector emanating from the origin -- the point representing the
identity element. The composition law for this group then 
corresponds to simply adding vectors representing the group elements  
using the parallelogram law.
 
Examined in detail, the parallelogram law for the translation group  
involves giving up  the picture in the last 
paragraph wherein each element of the  group was represented by a 
vector with tail  pegged to the origin, and going over to a picture based on 
free vectors -- 
vectors with their tails unpegged. Let $(\bx, \by)$ represent the free 
vector with head at $\by$ and tail at $\bx$. The group  element  
corresponding to translation by amount 
 $\ba$ is represented not by a single free vector, but by the {\em equivalence 
class} of all free vectors $(\bx,\by)$  obeying the only  condition             
$\by-\bx= \ba$. The parallelogram law for composing two group 
elements $(\bx, \by)$ and $({\bu},\bv)$ in 
this picture simply amounts to choosing representative free 
vectors, one  from either equivalence class, in such a manner that 
 the head of the first free vector  
coincides with  the tail of the second. We may use, for instance, the 
equivalence $(\bu,\bv) = (\by, \by + \bv - \bu)$. Then  the product of the 
two group elements is the free vector  
 from the tail of the first to the head of the second vector:
\begin{eqnarray}
(\bu,\bv)(\bx, \by) = (\by, \by +\bv - \bu)(\bx,\by) = (\bx,\by +\bv -\bu).
\end{eqnarray} 

Note that the equivalence class of free vectors corresponding to a group 
element  is obtained from the unique  vector
pegged to the origin by the left action of the translation group. Thus, the above 
construction 
can be carried over to any group $G$ through  left action of the group on
 the group manifold itself. More precisely, to each group element $g$ one can
associate an equivalence class of pairs $(g_0,gg_0)$, with the tail $g_0$
running over the entire group manifold. It can easily be verified that
whatever was said about the translation  group  goes
through here as well. In particular, composing two group elements 
$g,\,g'$  
represented by the equivalence classes corresponding respectively to $(g_0,gg_0)$ and 
$(g_{0}^{\prime}, g^{\prime}g_{0}^{\prime })$  requires us to choose the 
representative pair from each equivalence class in such a manner that the head $g 
g_0$  
 of the first  one coincides with   the tail of the second pair,   
$(g_{0}^{\prime}, g^{\prime}g_{0}^{\prime}) \sim 
( g g_{0},g^{\prime} g g_{0})$, 
 so that the resulting equivalence class
$(g_{0}, g^{\prime} g g_{0})$ corresponds to
the group element $g^\prime g$, thus endowing the composition law of
the arbitrary group $G$ with the structure of the parallelogram rule.             

This naive generalization of the parallelogram law applies uniformly to all 
groups, and  does not take advantage of  the specific features of a given 
group. It requires 
us to ascribe to each element of the group an equivalence class of pairs of points 
on the  group manifold, and this equivalence class is exactly as large as 
the group 
manifold itself. Thus, in the case of $SU(2)$ for which the group manifold is $S^3$ 
one would be associating with 
each group element  an equivalence class of pairs of points  on $S^3$ 
(the tail 
point of a pair can be considered to be arbitrary, and the 
head point is then fixed by the tail and the  group element under consideration). 
The importance of Hamilton's work lies in his recognizing that formulation of the 
$SU(2)$ composition law as a parallelogram rule can be accomplished,  with
greater economy, using equivalence classes of pairs of points on $S^2$ rather 
than on $S^3$.

The relationship between $SU(2)$ and $S^2$ is that the latter is an adjoint orbit 
of the former. That the parallelogram law for composition of $SU(2)$ group 
elements can be constructed on the (smaller) adjoint orbit without having to 
resort to the full group manifold rests ultimately on the fact that $SU(2)$ is 
strongly nonabelian (its centre is discrete). In comparison, since the 
translation group 
of the $n$-dimensional Euclidean space is abelian, parallelogram law 
for it cannot be constructed in any space smaller than the group manifold ${\cal 
R}^n$.

Just as $SU(2)$, the pseudo-unitary semisimple group $SU(1,1)$,  which 
is isomorphic  
to the real symplectic group $Sp(2,R)$ of linear canonical transformations, plays 
an important role in several areas of physics like squeezed light, Bogoliubov 
transformations, Gaussian or first order optics,  transmission lines, and reflection 
and transmission of classical 
and Schr\"{o}dinger waves at lossless boundaries and through barriers. Further, 
$SU(1,1)$ is the double cover of the (pseudo-orthogonal) Lorentz group $SO(2,1)$ 
 of  the $(2 + 1)$-dimensional space-time, a relationship similar to the one between 
$SU(2)$ and $SO(3)$.

Another low dimensional group whose importance for physics cannot be 
over-emphasised is the semisimple group $SL(2,C)$. This group is the double (and 
universal) cover of the Lorentz group $SO(3,1)$. Moreover, the complex orthogonal 
group $SO(3,C)$ is isomorphic to $SO(3,1)$: under a Lorentz transformation  the 
 space-time coordinates transform as an $SO(3,1)$ vector, but  the three 
 components 
 of $ \bE \pm i \bH$, where $\bE$ and $\bH$ are the electric and magnetic 
field vectors, transform as mutually conjugate   $SO(3,C)$  vectors.   

As noted above, Hamilton's  theory of turns has already been generalized to 
$SU(1,1)$. The purpose of the 
present paper is to develop a theory of turns for $SL(2,C)$ so that a geometric 
representation in the spirit of Hamilton will be available for all the low 
dimensional semisimple Lie groups of interest in physics.

The contents of this paper are organized as follows. In Section 2 we recount 
briefly Hamilton's theory of turns for $SU(2)$, and also its  generalization 
to $SU(1,1)$. This summary should prove useful in view of the fact that the theory of 
turns  that we develop for $SL(2,C)$  runs almost  parallel to that in the above 
cases. The adjoint 
orbits in the Lie algebra of $SL(2,C)$ are considered in Section 3, and the 
relationships between the groups $SO(3,1),\; SO(3,C)$  and $SL(2,C)$ are indicated. 
In Section 4 we construct turns for $SL(2,C)$ in the adjoint orbit of complex 
``unit'' vectors $ \Sigma = \{\hz = (z_1,  z_2, z_3), \;  z_1\,^2 + z_2\,^2 + 
z_3\,^2 = 1\}$, 
and demonstrate the parallelogram law for composition of turns. As an exercise in 
the use of turns,  in Section 5 we describe the polar decomposition in the 
language of turns. Composition of Lorentz boosts and the resulting Wigner (Thomas) 
rotation are studied in Section 6 using turns. We conclude in Section 7 with some 
final remarks.

\section { Turns for $SU(2)$ and $SU(1,1)$}

In the defining $2 \times 2$ representation, elements of the 
group $SU(2)$ are described  in terms of the Pauli matrices $\bs = (\sigma_1,\, 
\sigma_2,\, \sigma_3)$ as 
\begin{eqnarray}
g(a_0, \ba) = a_0  - i\ba\cdot \bs\,,
\end{eqnarray} 
where $a_0$ is a real scalar and $\ba$ is a real three vector
satisfying the constraint
\begin{eqnarray}
a_{0}^{2} + \ba\cdot\ba =1\,.
\end{eqnarray}
Thus the group manifold of $SU(2)$ is the unit sphere 
$S^3$. 
 The centre of $SU(2)$ is $Z_2$, the subgroup consisting of the two elements 
$\pm 1$.

The Lie algebra  considered as a linear space coincides with 
${\cal R}^3$, and it consists of all traceless  hermitian matrices  $\bx 
\cdot \bs,\;\; \bx \in {\cal R}^3$. 
The group $SU(2)$, modulo its center $Z_2$,  can be realized by its adjoint action 
on its Lie algebra:
\begin{eqnarray}  
 g:\;\; \bx \cdot \bs \to  g\, \bx\cdot\bs\, g^{-1} &=& 
\bx ^{\prime}\cdot \bs\, , \nonumber\\
\bx^\prime = R(g) \bx,\;\; R(g) &\in& SO(3) = SU(2)/ Z_2\,.
\end{eqnarray}
Thus $\bx \cdot \bx = -\det \bx \cdot \bs$ is invariant, and hence  the 
adjoint orbits  of 
$SU(2)$ are spheres centred at the origin. Hamilton's turns can be constructed on 
any of these orbits, and we choose the unit sphere  $S^2$.

The construction goes as follows. Given $g(a_0, \ba) \in SU(2)$, choose a pair of 
unit vectors $\hx, \hy \in  S^2$ such that  
$a_0 = \hx.\hy, \;\ba = \hx \wedge \hy$. This is always possible and, indeed, 
there exists not just one choice but an equivalence class of choices.  
When $\ba \ne  0$, the unit   vectors $\hx, \hy$  
are necessarily  orthogonal to $\ba$, and there exists a one parameter worth of 
freedom in 
the choice of $\hx, \hy$: for instance, $\hx$ can be chosen to be an arbitrary 
point on the great circle of $S^2$ perpendicular to $\ba$, and then $\hy$ is 
uniquely determined by $\hx \wedge\hy = \ba$ and $\hx \cdot \hy = a_0$. On the 
other hand,  when ${\ba } = 0$, i.e. when $g(a_0, \ba) = 
\pm 1$, the vectors $\hx, \hy$ are necessarily  parallel or antiparallel, and there 
is a two 
parameter worth  of freedom in choosing the pair ( $\hx$ may be chosen to be an 
arbitrary point,  and then $\hy$ is fixed: $(\hx,\,\hy) = (\hx, \pm \,\hx)$). 

We may use $\hx, \hy$ to 
label the elements of $SU(2)$ and write
\begin{eqnarray}
g(\hx, \hy) = \hx \cdot \hy - i \hx \wedge \hy \cdot \bs\,.
\end{eqnarray}
This equation describes  a one-to-one correspondence between elements of $SU(2)$ and 
equivalence classes of pairs of points in $S^2$. 

Now, a pair of points in $S^2$ is 
the same thing as a directed great circle arc with tail at the first member of the 
pair and head at the second. Thus emerges the one-to-one correspondence between 
elements of $SU(2)$ and equivalence classes of directed great circle (geodesic) arcs. 
{\it These 
equivalence classes are the turns of Hamilton}. Arcs of an equivalence class belong 
to the same great circle, have the same sense and same arc length $\le \pi$, so 
that the members of an equivalence class are  obtained by rigidly  sliding 
one arc of the class  along 
its own geodesic. The element $- 1 \in SU(2)$ corresponds to the 
equivalence class 
of great semicircles or antipodal points, and the identity element to the equivalence 
class of null arcs or singleton points.

Since $g(a_0,\ba) ^{-1}=g(a_0,\,-\ba)$, we have $g(\hx,\hy)^{-1} = 
g(\hy,\hx)$. That 
is,  the $SU(2)$ inverse of a turn corresponds simply  to 
reversal of the sense of the turn. The group composition law rests on the identity
\begin{eqnarray}
g(\hy,\hz) g(\hx,\hy) = g(\hx,\hz)\,,
\end{eqnarray}
which follows from the definition (2.4). This is the parallelogram law for the 
product of two 
$SU(2)$ elements when the tail of the left  factor coincides with ``and 
cancels''  the head of the 
right factor at $\hy$. The product then  corresponds to the directed great circle arc 
from 
the free tail $\hx$ to the free head $\hz$, as shown in Fig.1.

That it is always possible, given two 
arbitrary elements of $SU(2)$, to choose directed arcs from the respective  turns so 
that the tail of the left element coincides with the head of the right 
element is 
guaranteed by the fact that great circles on $S^2$ intersect (or coincide). Thus,  
the parallelogram law (2.5) faithfully reproduces $SU(2)$ multiplication 
(including its noncommutativity!).

\begin{figure}[h]
\epsfxsize8cm
\centerline{\epsfbox[140 340 420 710]{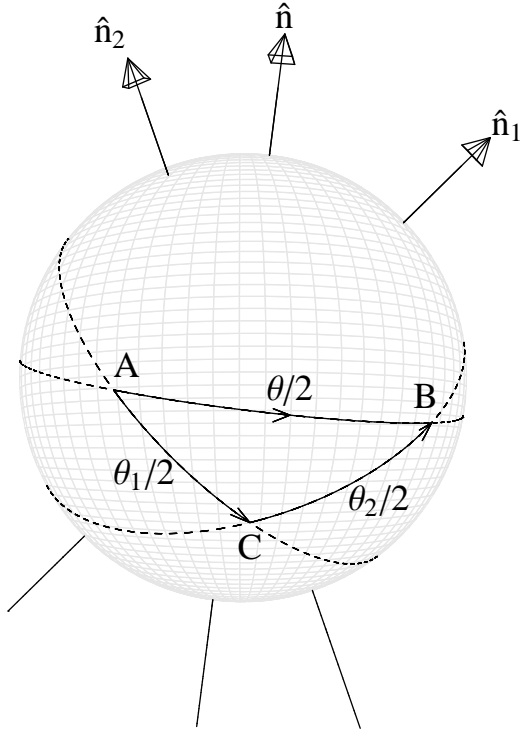}}
\caption{Showing the parallelogram law of group composition for $SU(2)$. The turns
representing $S(\theta_1 ,\, \protect\hn_1)$ and $S(\theta_2 ,\, \protect\hn_2)$ live 
in the
equators orthogonal to $\protect\hn_1$ and $\protect\hn_2$ respectively. The head of 
$S(\theta_1
,\, \protect\hn_1)$ and the tail of $S(\theta_2 ,\, \protect\hn_2)$ meet at C.  The 
directed
geodesic arc AB from the free tail of $S(\theta_1 ,\, \protect\hn_1)$ to the free 
head of $S(\theta_2 ,\, \protect\hn_2)$ represents the turn corresponding to the 
product
$S(\theta ,\, \protect\hn ) = S(\theta_2 ,\, \protect\hn_2) S(\theta_1 ,\, 
\protect\hn _1)$, with 
$\protect\hn$ orthogonal to the geodesic AB and $\theta =$ twice the arclength of AB.}
\end{figure}

\noindent
Given $a_0 - i \ba \cdot \bs \in SU(2)$, the condition $a_0\,^2 + \ba \cdot \ba = 
1$  implies that we can find a unit vector $\hn$ and an angle $ -2\pi< 
\theta\leq2\pi$ such that $a_0 = \cos \theta /2$ and  $ \ba = \hn \,\sin \theta 
/2\,$. Thus, we may also parametrize the elements of $SU(2)$ as 
\begin{eqnarray*}
S(\theta ,\, \hn)
\,= \, \cos  \theta /2 \, - \,i\,\hn\cdot\bs \,\sin \theta /2\,,
\end{eqnarray*}
rendering the fact  that $S^3$, the group manifold of $SU(2)$, is a $U(1)$ bundle 
over $S^2$. In the adjoint representation $S(\theta ,\, \hn)$ acts as rotation 
through angle $\theta$ about the direction $\hn$. While  the  unit vectors 
$(\hx,\,\hy)$ in (2.4) live in the ``equator'' orthogonal to $\hn$, the angle between 
them is $\theta /2$ and not $\theta$ (see Fig. 1).

This concludes our account of Hamilton's theory of turns for $SU(2)$. We have gone 
over it in some detail, for the  theory of turns that we shall develop for the 
Lorentz group will follow the $SU(2)$ case quite closely.

The theory of turns has turned out to be a powerful tool in connexion with several 
synthesis problems in polarization optics\cite{pramana}. A particularly striking 
result  in 
this context is the following: {\it  every intensity preserving} [i.e., $SU(2)$] 
{\it transformation in polarization optics can be synthesised with just two quarter 
wave plates and one half wave plate}\cite{pramana,gadget1,gadget2,gori}. Moreover, an 
intimate connection between 
$SU(2)$ turns and the Berry-Pancharatnam\cite{geophase} geometric phase has  been 
established\cite{jphys}.  

Hamilton's theory of turns has been generalized to the noncompact  group 
$SU(1,1)$ only recently\cite{prl,jmp}. This group  consists of two dimensional 
complex pseudounitary matrices of unit determinant: 
\begin{eqnarray}
g = \left(\begin{array}{cc} \alpha&\beta\\ \beta^* & \alpha^* \end{array}
\right),\;\;\; {\mid \alpha \mid}^2 - {\mid \beta \mid}^2 =1\,.  
\end{eqnarray} 
It order to maintain a close similarity with the
group $SU(2)$, it proves convenient to define
\begin{eqnarray}
\br &=& (\rho_1, \rho_2, \rho_3)\,,\nonumber\\
\rho_1 = \sigma_3\,,~~ \rho_2 &=& i\sigma_1\,,~~ \rho_3 = i\sigma_2\,,
\end{eqnarray} 
and to introduce the $(2+1)$-Minkowskian analogues (with indices running over 
1,2,3 rather than over 0,1,2) of the scalar and the cross products of vectors as 
follows \begin{eqnarray}
\bx\cdot\by \equiv  \eta_{ab}x^a y^b ,~~~ \bx \wedge
\by \equiv {\epsilon^a}_{bc} x^b y^c,
\end{eqnarray}
where $\eta_{ab} = \mbox{diag}(-1,+1,+1)$ is the metric,  and
$\epsilon_{abc}$ is the Levi-Civita symbol with $\epsilon_{123} = 1$. With 
the help of this notation $SU(1,1)$ matrices, in analogy with the $SU(2)$
case,  can be expressed as
\begin{eqnarray}
g(b_0, \bb) = b_0 - i\bb \cdot \br\,,
\end{eqnarray} 
where the scalar $b_0$ and the vector $\bb$ are real and are constrained by
\begin{eqnarray}
b_0\,^2 -\bb\cdot \bb = b_0\,^2 + b_1\,^2 - b_2\,^2 - b_3\,^2  =1\,.
\end{eqnarray}
The group manifold of $SU(1,1)$ is the hyperboloid described by (2.10). 
Topologically it has the same structure as ${\cal R}^2 \times S^1$, as can be seen 
by solving for $b_0$ and $b_1$ in (2.10) in terms of $(b_2, b_3) \in {\cal R}^2$ 
and an angle variable $\theta$: 
\begin{eqnarray}
b_0 &=& \sqrt{1 +  b_2\,^2 + b_3\,^2}\, \cos \theta\,, \nonumber\\
b_1 &=& \sqrt{1 +  b_2\,^2 + b_3\,^2}\, \sin \theta\,. 
\end{eqnarray}
Thus, the group manifold of $SU(1,1)$ is noncompact. It is  multiply connected, while 
that of the compact group $SU(2)$ is simply connected.
 However, the centre of $SU(1,1)$ is $Z_2$, the subgroup consisting of the two 
elements $\pm 1$, just as  with $SU(2)$.

The Lie algebra of $SU(1,1)$  
 consists of the matrices  $\bx \cdot \br,\;\; 
\bx \in {\cal R}^3$. 
  Adjoint action of $SU(1,1)$
on its Lie algebra gives a realization of $SU(1,1)$ modulo its center $Z_2$: 
\begin{eqnarray}  
 g:\;\; \bx \cdot \br \to  g\, \bx\cdot\br\, g^{-1} &=& 
{\bx}^{\prime}\cdot{\br}\,, \nonumber\\
\bx^\prime &\equiv&\Lambda (g) \bx,\;\;\nonumber\\
 \Lambda(g) \in SO(2,1) &=& SU(1,1)/ 
Z_2\,. 
\end{eqnarray}
We made use of the fact that  $ \det \bx \cdot \br = \bx \cdot \bx = x_2\,^2 + 
x_3\,^2 - x_1\,^2$ is invariant under conjugation

Whereas all the adjoint orbits of $SU(2)$ were of the same type, namely concentric 
spheres, in the $SU(1,1)$ case there are different types of orbits since, 
 owing to the indefinite 
nature of the Minkowskian scalar product, the invariant $\bx \cdot \bx = \det 
\bx \cdot \br$ can be 
positive, negative, or zero (adjoint orbits of the orthogonal and 
pseudo-orthogonal groups $SO(m,n),\; m+n \leq 5$, have been analysed and classified 
in Ref.\cite{orbits}). 
 The single sheeted unit hyperboloid  consisting  of spacelike
unit vectors $\bx$ with $ x_2\,^2 + x_3\,^2 - x_1\,^2 = 1$ proves 
convenient\cite{prl,jmp} in generalizing Hamilton's theory of turns to 
$SU(1,1)$. Indeed, the 
construction proceeds in close analogy with the $SU(2)$ case, and the details can 
be found in Refs.\cite{prl,jmp}.

\section{ $SL(2,C)$ and its connection with $SO(3,C)$ and $SO(3,1)$}
The group $SL(2,C)$ in its defining representation consists of complex 
valued $2 \times 2$ matrices of unit determinant: 
\begin{eqnarray}
S = \left(\begin{array}{cc} a&b\\c & d \end{array}\right),\;\; ad - bc = 1\,.
\end{eqnarray} 
It is a noncompact six parameter real Lie 
group. The centre of $SL(2,C)$ is $Z_2 = \{\pm 1\} $, and the group manifold is 
simply connected.  
 
The Lie algebra consists of all  traceless complex  matrices $\bz \cdot \bs,\; 
\bz \in {\cal C}^3$. In other words  the one parameter subgroups are of the form 
\begin{eqnarray} 
g_{\bz}(t) = \exp (-it\bz\cdot\bs) &=&
\exp [-it(\bx\cdot\bs + i\by\cdot\bs)]\,, \nonumber\\
\bx, \by &\in& {\cal R}^3, \;\;\bz = \bx+i\by \in {\cal C}^3. 
\end{eqnarray}

The connection between $SL(2,C)$ and $SO(3,C)$ is exposed  by the adjoint action  
of the former on its Lie algebra:
\begin{eqnarray}  
 S:\;\; \bz \cdot \bs \to  S\, \bz\cdot\bs\, S^{-1} &=& 
{\bz}^{\prime}\cdot{\bs}\,, \nonumber\\
\bz^\prime = R^c  (S) \bz,\;\; R^c (S) &\in& SO(3,C) = SL(2,C)/ Z_2\,.
\end{eqnarray}
Here we made use of the fact that the trace and the determinant of a matrix are 
invariant under conjugation, and the fact that $\det \bz \cdot \bs = - \bz \cdot 
\bz$. The superscript over $R$ is to remind us that we are dealing with the 
complex, rather than the real, orthogonal group, and $\bz$ is to be viewed as a 
column vector.

For any Lie group $G$ with centre $\Omega$, the adjoint representation is a faithful 
representation of the quotient group $G/\Omega$. In the cases of $SU(2)$ and 
$SU(1,1)$ 
 we have seen that the adjoint representations coincide respectively with the 
defining 
representations of $SO(3)$ and $SO(2,1)$, consistent with the fact that 
$SU(2)/Z_2 = 
SO(3),\;\;SU(1,1)/Z_2 = SO(2,1)$. In the case of $SL(2,C)$ we find that the adjoint 
representation coincides with the defining representation of 
 the complex orthogonal group  $SO(3,C)$, with the complex $3\times 3$ matrices of 
the latter rewritten as real $6\times 6$ matrices in a natural manner.

The group $SL(2,C)$ is related  also  to the Lorentz group  $SO(3,1)$ in  
a well known two-to-one manner. To see this, consider a generic hermitian matrix 
\begin{eqnarray}
H = \xi_0 + \bxi\cdot\bs\,,
\end{eqnarray}
where $\xi_0$ is a real scalar and $\bxi \in {\cal R}^3$. We have 
 $\mbox{tr} H = 2\xi_0$ and $\det H = \xi_{0}\,^{2} - 
\bxi\cdot\bxi$.  Clearly, for any $S \in SL(2,C)$, the map $S:\; H \rightarrow 
SHS^\dagger$ preserves $\det H$, hermiticity of $H$, and the signature of 
$\mbox{tr} H$ (the numerical 
value of $\mbox{tr} H$ is not preserved since the map is a congruence and not 
conjugation): 
\begin{eqnarray} 
S:~~ \xi_0 + \bxi \cdot \bs &\to& S(\xi_0 +\bxi\cdot\bs )S^\dagger = 
\xi_{0}^{\prime} +\bxi^{\prime}\cdot\bs\,, \nonumber\\
\xi_{0}^{\prime 2} - \bxi^\prime \cdot\bxi^\prime &=& \xi_{0}^{2} - 
\bxi\cdot\bxi\,, \nonumber \\
\mbox{sign} \,\xi_{0}^\prime &=& \mbox{sign}\, \xi_0\,. 
\end{eqnarray}
This implies that $(\xi_0',\bxi ')$ and $(\xi_0, \bxi)$ are related by a  real 
$4\times 4$ Lorentz transformation matrix $L(S)$: 
\begin{eqnarray}
\left(\begin{array}{c} \xi _{0}^\prime \\\bxi ^\prime \end{array}\right)
= L(S) \left(\begin{array}{c} \xi _{0}\\\bxi \end{array}\right),\; \;\; L(S) \in 
SO(3,1)\,. 
\end{eqnarray}
 Since $Z_2 = \{ \pm 1 \}$ is the kernel of the map (3.5), one concludes that  the 
proper 
orthochronous Lorentz group is isomorphic to $SL(2,C)/Z_2$:
\begin{eqnarray}
 SO(3,1) \sim SL(2,C)/Z_2 \sim SO(3,C)\,.
\end{eqnarray}

We may illustrate the connections between these three groups by considering some one 
parameter subgroups (OPS's). The compact OPS $\exp(-i\frac{\theta} {2} \sigma _1)$ 
generated by (the hermitian) $\sigma _1$ takes the following forms    
\begin{eqnarray}
g(\theta) &=& \left(\begin{array}{cc} 
\cos (\theta/2)&-i\sin (\theta/2)\\ 
-i\sin (\theta/2)&\cos (\theta/2)
\end{array} \right),\;\;\; 0\le\theta<4\pi\,;\nonumber\\  
R^c(\theta) &=& \left(\begin{array}{ccc}
1&0&0\\ 
0&\cos \theta&-\sin\theta\\ 
0&\sin \theta&\cos\theta
\end{array} \right),\;\;\; 0\le\theta<2\pi\,;\nonumber\\  
\Lambda(\theta) &=& \left(\begin{array}{cccc}
1&0&0&0\\
0&1&0&0\\ 
0&0&\cos \theta&-\sin\theta\\ 
0&0&\sin \theta&\cos\theta
\end{array} \right),\;\;\; 0\le\theta<2\pi\,;  
\end{eqnarray} 
respectively in $SL(2,C),\; SO(3,C)$ and $SO(3,1)$. On the other hand, the 
noncompact OPS $\exp(\frac{\beta} {2} \sigma _1)$ 
generated by (the antihermitian) $i \sigma _1$ takes the following forms:    
\begin{eqnarray}
g(\beta) &=& \left(\begin{array}{cc} 
\cosh (\beta/2)&\sinh (\beta/2)\\ 
\sinh (\beta/2)&\cosh (\beta/2)
\end{array} \right),\;\;\; \beta \in {\cal R}\,;\nonumber\\  
R^c(\beta) &=& \left(\begin{array}{ccc}
1&0&0\\ 
0&\cosh \beta&-i\sinh\beta\\ 
0&i\sinh \beta&\cosh\beta
\end{array} \right),\;\;\; \beta \in {\cal R}\,;\nonumber\\  
\Lambda(\beta) &=& \left(\begin{array}{cccc}
\cosh\beta&\sinh\beta&0&0\\
\sinh\beta&\cosh\beta&0&0\\ 
0&0&1&0\\ 
0&0&0&1
\end{array} \right),\;\;\; \beta \in {\cal R}\,. 
\end{eqnarray} 
It is the identification of $\theta$ and $\theta + 2\pi$ for compact OPS's in 
going over from $SL(2,C)$  to $SO(3,C)$ or $SO(3,1)$ that is ultimately 
responsible for the fact that the latter groups are topologically different from 
the former.

The adjoint orbits\cite{lorentz,orbits} of the Lie algebra of $SL(2,C)$  can be 
easily figured out from the fact that  $\bz\cdot\bz$
 is  the only (complex) invariant associated with a generic  Lie algebra element 
$\bz \cdot \bs$ under the adjoint 
action. Since   $\bz\cdot\bz$   can assume any numerical  
value, it is clear that there exists precisely one orbit for each point in the 
complex plane. Closer examination shows that these orbits fall into two different 
types\cite{lorentz,orbits}.\\

\noindent
\underline{{\bf type-I orbits}}: $\bz\cdot\bz = r^2 \mbox{e}^{2i\phi},\;0\le \phi< 
\pi,\;0<r< \infty $\,.\\

\noindent
In this case the  Lie algebra element $\bz \cdot \bs$ with $\bz \cdot \bz 
\neq 0$ can be brought, using the adjoint action, into the  canonical form 
\begin{eqnarray}
\bz = (z_1, z_2, z_3) &\to& \bz _0 = (r\mbox{e}^{i\phi}, 0,0)\,,\nonumber\\
\bz \cdot \bs &\to& \bz _0 \cdot \bs = r \mbox{e}^{i\phi}\,\sigma_1\,.
\end{eqnarray}
Clearly, the stability group of this canonical form is
\begin{eqnarray}
G_I&=& SO(2) \times SO(1,1) = SO(2,C) \nonumber\\
&=&\left\{ \exp 
\left(-\frac{i}{2}(\alpha + i 
\beta)\sigma _1\right),\;\; \; 0\le\alpha<4\pi,\;\; \beta \in {\cal R}\right\},
\end{eqnarray}
which has the topology of the cylinder ${\cal R} \times S^1$. Thus the type-I 
orbits have the structure of the four dimensional manifold  $SO(3, C)/G_I$, and there 
exists precisely  one such orbit for every nonzero complex number.\\

\noindent
\underline{{\bf type-II orbit}}: $\bz\cdot\bz = 0$\,.\\

\noindent
A  Lie algebra element $\bz \cdot \bs$ with $\bz \cdot \bz 
= 0$ can be brought, using the adjoint action, into the  canonical form 
\begin{eqnarray}
\bz = (z_1, z_2, z_3) &\to& \bz _0 = (1,i,0)\,,\nonumber\\
\bz \cdot \bs &\to& \bz _0 \cdot \bs = \sigma _1 + i \sigma_2\,.
\end{eqnarray}
The stability group of this canonical form is
\begin{eqnarray}
G_{II} = \left\{ \exp \left(-\frac{i}{2}(\alpha + i \beta)(\sigma_1 + i\sigma 
_2)\right),\;\; \; \alpha,\beta \in {\cal R} \right\},
\end{eqnarray}
which has the topology of  ${\cal R}^2$. Thus the type-II 
orbit has the structure of the four dimensional manifold  $SO(3, C)/G_{II}$. Both 
$G_I$ and $G_{II}$ are abelian. At 
the risk of repetition we wish to note that there exists only one type-II orbit.

To construct turns for $SL(2,C)$ we shall make use of the orbit $ \bz\cdot
\bz =1$. We shall denote this orbit by $\Sigma$, and refer to it as  {\it the 
orbit of complex unit vectors}. For convenience, elements of $\Sigma$ will be 
decorated with a hat, like $\hx,\hy,\ldots$

To conclude this Section, we note that the vector algebra in ${\cal R}^3 $  
carries over to $C^3$ with virtually no change. In particular, the following  
familiar identities are true for complex-valued vectors as well:
  \begin{eqnarray}
\ba\wedge (\bb\wedge \bc)
&=&\bb(\ba\cdot \bc)- 
\bc(\ba\cdot \bb)\,,\nonumber\\
(\ba\wedge \bb)\cdot (\bc\wedge {\bd}) &=& 
(\ba\cdot \bc)(\bb\cdot {\bd})-
(\ba\cdot {\bd}) (\bb\cdot \bc)\,,\nonumber\\
(\bz\cdot \by)(\bx\cdot {\bz }) &-& 
(\bz\wedge \by)\cdot(\bx\wedge \bz)=
(\bz\cdot \bz)(\bx\cdot \by)\,,\nonumber\\
(\bz\cdot \bz)(\bx\wedge \by)
&=&(\bz\cdot \by)(\bx\wedge \bz)+
(\bz\cdot \bx)(\bz\wedge \by)\nonumber\\ 
&& \;\;\;\;\;\;\;\;+ (\bz\wedge \by)\wedge (\bx\wedge \bz)\,.
\end{eqnarray}
These relations will be found useful in the next Section.

\section{Construction of turns for $SL(2,C)$}
We now have all the ingredients to construct a theory of turns for  $SL(2,C)$.  
Elements of $SL(2,C)$   can be described through 
\begin{eqnarray}
S(a_0, \ba) = a_0 -i\ba \cdot\bs\,,
\end{eqnarray}
where  $a_0$ is a complex scalar and $\ba$ a complex three vector subject to the 
condition  
 $a_{0}^2 + \ba\cdot\ba =1$. This is  one complex condition on four complex 
numbers. Given $S(a_0, \ba) \in SL(2,C) $, to construct the turn corresponding to 
$S(a_0,\ba)$  we look for  pairs of 
complex unit  vectors $\hx, \hy$ such that $a_0 = \hx \cdot \hy$ and $\ba = \hx 
\wedge \hy$.  

Let $\hx$ be a unit 
vector perpendicular to $\ba$,  i.e. $\hx\cdot\ba= 0$. This corresponds to two real 
constraints in the four-parameter manifold of unit vectors. Thus, there is a 
two parameter worth of freedom in the choice of $\hx$ perpendicular to 
$\ba$. Define  
\begin{eqnarray}
\by \equiv  a_0 \hx + \ba\wedge \hx\,.
\end{eqnarray}
Then  $a_{0}^2 + \ba\cdot\ba =1$ guarantees that $\by\cdot \by 
= 1$, and we may indeed  write $\hy$ in place of $\by$. Further, it can be verified  
that the conditions  $a_0 =  
\hx\cdot \hy$ and $\ba=\hx\wedge \hy$ are satisfied. 
Since there is a 
two-parameter worth of  freedom in the choice of $\hx$, we have not just one pair 
but  an equivalence 
class of pairs $\hx,\hy$ representing a given $S \in SL(2,C)$. 
 We may use 
any one of 
these pairs and  write \begin{eqnarray} S(a_0, \ba) &=& a_0 -i\ba.\bs \nonumber \\
 &=& \hx\cdot \hy -i\hx\wedge \hy\cdot \bs
\equiv  S(\hx, \hy)\,,
\end{eqnarray} 
and call the equivalence class  $S(\hx, \hy)$ as the turn from the tail $\hx$ to the 
head $\hy$. (The improper  use of a common symbol $S(\,\cdot \,,\,\cdot \,)$ at the 
start and the end of this equation  should cause no confusion.) 
 Conversely, for any choice of two complex unit vectors $\hx$ and $\hy$, $S(\hx, 
\hy)$ is indeed an  $SL(2,C)$ element, and two pairs $\hx, 
\hy$ and $\hx ',\hy '$ are in the same equivalence class if and only if $\hx 
\cdot \hy = \hx '\cdot  \hy '$ and $\hx \wedge \hy = \hx ' \wedge \hy '$. 
For the special elements $\pm 1 \in SL(2,C)$, we have $\ba = 0$ and 
hence $\hy = \pm \hx$.

 Since  $S(a_0, \ba)^{-1} = S(a_0, -\ba)$,  it follows that 
\begin{eqnarray}
{S(\hx, \hy)}^{-1} =S(\hy, \hx)\,.
\end{eqnarray}
Thus the inverse  corresponds to reversing the sense of the turn, as one would 
have wished. Only the parallelogram law of composition remains to be established.

With the aid of the identities (3.14), it is readily seen that
  the definition (4.3) possesses the following property: 
\begin{eqnarray}
S(\hz, \hy)~S(\hx, \hz)=S(\hx, \hy)\,.
\end{eqnarray} 
This is the parallelogram law when the head of the right factor coincides with 
the tail of the left one. The  theory of turns for $SL(2,C)$ will be deemed  
complete if one can show 
that there always exists a common unit vector $\hz$ where two turns meet.

Suppose $a_0 -i\ba.\bs$ and $b_0 -i\bb.\bs$ are 
two $SL(2,C)$ elements, then  the common unit vector $\hz$ where the corresponding 
turns meet should necessarily  satisfy  $\ba.\hz=0$ and $\bb.\hz=0$. Thus $\hz$ must 
be a 
multiple of $\ba\wedge\bb$. It follows that  a unit vector $\hz$ common to two 
turns exists if and only if  
$(\ba\wedge \bb)\cdot(\ba\wedge \bb)\neq 0 $. 

Thus, if the two $SL(2,C)$ elements to be composed are such that 
$(\ba \wedge \bb) \cdot ( \ba \wedge \bb) \ne 0$, 
then their turns meet in the space of unit 
vectors $\Sigma$, and (4.5) constitutes the parallelogram law of composition for such
 pairs. But if $(\ba \wedge \bb) \cdot ( \ba \wedge \bb) = 0$, the corresponding 
turns do not meet in $\Sigma$, and hence further work is needed before (4.5) can be 
used as the parallelogram law of composition for such pairs. 
 Fortunately,  the latter situation occurs  only  in a measure zero set of cases: 
$(\ba\wedge \bb)\cdot(\ba\wedge \bb)$ can take any complex  value, but the 
turns fail to meet in $\Sigma$  only when this value is zero.  

In such special cases one may infinitesimally modify one of the two $SL(2,C)$ 
elements to be composed, say change $S(b_0, \bb)$  to $S(b_0', \bb')$ 
 so that $b_0' - b_0$ and $\bb' -\bb$ are infinitesimals and 
$(\ba \wedge \bb') \cdot ( \ba \wedge \bb') \ne 0$, use (4.5) to compose the 
turns, and take the limit of the composed turn as the infinitesimals go to zero. 
Alternatively, we may suitably factorize $S(b_0, \bb)$ into a product of two 
$SL(2, C)$ elements, $S(b_0,\bb) = S(b_0'', \bb'')S(b_0' , \bb')$, and use 
the parallelogram law twice: compose the turn corresponding to $S(a_0,\ba)$ with 
that corresponding to $S(b_0', \bb')$, and then compose the resulting turn with 
that corresponding to $S(b_0'',\bb'')$ (this latter approach is the one used 
  in the theory of turns for $SU(1,1)$ presented in Ref.\cite{prl,jmp} for handling 
turns which do not meet).

With these provisions for handling the measure zero case of nonintersecting turns, 
(4.5) constitutes the parallelogram law of composition for $SL(2,C)$ 
turns.

\section{ Turns and the Polar Decomposition}
In this Section we describe in the language of  turns we have just developed the 
 process of polar decomposition. Such an exercise not only helps us in developing 
a feel for $SL(2,C)$ turns,  but also helps in the study  of composition of 
Lorentz boosts to be taken up in the next Section.

Any element  $ S \in SL(2,C)$ can be decomposed as $S=HU$, where $H$ is hermitian  
positive 
definite, $U$ is unitary, and therefore  $H, U \in SL(2,C)$. This polar  
decomposition is unique for a given $S$, and  corresponds to the decomposition 
of an element $\Lambda \in SO(3,1)$ into a spatial rotation followed by a 
 boost. That is,
 $\Lambda = PR$ where $P$ is real symmetric positive definite and $R$ is 
orthogonal. Again, $P$ and $R$ are uniquely determined in terms of $\Lambda$, and 
both factors are elements of $SO(3,1)$.  In $SO(3,C)$, this decomposition 
corresponds to a rotation through a purely real angle followed by a rotation
through a purely imaginary angle.  It is important to appreciate that polar 
decomposition is not covariant under conjugation by  
$SL(2,C)$ [equivalently, by $SO(3,1)$ or $ SO(3,C)$]. It is  covariant only under 
$SU(2)$ [equivalently, under  $SO(3)$].
   
Now an element $a_0 -i\ba.\bs \in SL(2,C)$ is unitary (real rotation) if and 
only if $a_0$ and $\ba$ are real. In the language of turns this translates into 
the following statement:  $S=\hx\cdot \hy -i(\hx\wedge \hy)\cdot \bs$ is unitary 
 if and only if $\hx$ and $\hy$ are real. Similarly $S= 
a_0 -i\ba.\bs$ is hermitian positive definite (pure boost) if and only 
if $a_0$ is real and $>0$ and $\ba$ is imaginary. In terms of turns this 
happens if  one member of the pair  
$(\hx$,$\hy)$, say $\hx$, is purely real to which the real part of 
the other, say $\hy$, is parallel (not antiparallel). The imaginary part of $\hy$ 
is then perpendicular to  $\hx$ by virtue of the fact that the real and imaginary 
parts of a complex unit vector are necessarily orthogonal.

Writing $\ba = \ba _R + i \ba _I$,  polar decomposition of $S(a_0,\ba) = a_0 
-i\ba.\bs$ is trivial when $ 
\ba _R$ and $\ba _I$,  the real and imaginary parts of $\ba$, are multiples of one 
another. So we assume that $\ba _R$ and $\ba _I$  are linearly independent. Our 
aim is to decompose $S(\hx, \hy) \in SL(2,C)$ in the polar form
\begin{eqnarray}
S &=& a_0 -i\ba.\bs \nonumber\\
 &=& S(\hx, \hy) = 
S(\hz, \hy)S(\hx, \hz)\,,
\end{eqnarray}
with $\hx,\hz$ real so that $S(\hx,\hz)$ will be an element of $SU(2)$ (i.e., 
rotation)  and the real part of $\hy$   parallel to $\hz$ so that $S(\hz,\hy)$ will 
be hermitian positive definite (i.e., pure boost).

We noted in Section IV that there exists a two parameter worth of freedom in 
choosing the tail point $\hx$ of a turn $S(\hx,\hy)$. To facilitate 
polar decomposition we need to choose $\hx$ to be real, and the above 
 two parameter freedom permits such a choice. Indeed, reality of $\hx$ and the 
requirement $\ba \cdot \hx = 0$, which follows from $\ba = \hx \wedge 
\hy$, together imply that $\hx$ is necessarily orthogonal to both the 
real and imaginary parts of $\ba = \ba _R + i \ba _I$:
\begin{eqnarray} \hx = \pm 
\frac{\ba_R \wedge \ba_I} {\sqrt{(\ba_R \wedge \ba_I)\cdot(\ba_R \wedge \ba_I)}}\,.
\end{eqnarray}
Then, from (4.2), 
\begin{eqnarray}
\hy &=& a_0 \hx + \ba\wedge \hx \nonumber\\
&=& a_{0R} \hx + \ba_{R}\wedge \hx
+i a_{0I} \hx + i \ba_{I}\wedge \hx\,,
\end{eqnarray}
where $a_{0R}$ and $a_{0I}$ are respectively the real and imaginary parts of the 
scalar $a_0$. Since $\hz$ has to be real and  parallel ( not antiparallel) to the 
real part of $\hy$, we have 
\begin{eqnarray}
\hz =  \frac {a_{0R} \hx + \ba_R \wedge \hx}
{\sqrt{a_{0R}^{2} +\ba_R \cdot \ba_R}}\,\,,
\end{eqnarray}
and the polar decomposition is completed. 

Reality of $\hx$ required by polar decomposition removed the two parameter worth 
of arbitrariness or freedom  in the choice of $\hx$, and hence in that of $\hz$ and $ 
\hy$, 
except for a signature in (5.2). Since $\hy$ and $\hz$   in (5.3), (5.4) are 
linear in $\hx$, and since the quantities entering the polar decomposition are  
quadratic in $\hx, \hy$ and $\hz$, either choice for the signature in (5.2) leads 
to the same set of expressions, confirming the uniqueness of the polar decomposition:
 \begin{eqnarray}
S(a_0, \ba) &=& a_0 - i \ba \cdot \bs \nonumber\\
            &=& S(\hz, \hy) S(\hx, \hz) \nonumber\\
	    &=& (\hz \cdot \hy - i \hz \wedge \hy \cdot \bs)
	     (\hx \cdot \hz - i \hx \wedge \hz \cdot \bs)\,;\nonumber\\
\hx\cdot\hz &=&  \frac {a_{0R}}
{\sqrt{(a_{0R})^{2} +\ba_R \cdot \ba_R}}\,\,, \nonumber\\
\hx\wedge \hz &=&  \frac { \ba_R }
{\sqrt{(a_{0R})^{2} +\ba_R \cdot \ba_R}}\,\,;\nonumber\\
\hz\cdot\hy &=&  \sqrt{(a_{0R})^{2} +\ba_R \cdot \ba_R}~~ >~~ 1\,,\nonumber\\
\hz\wedge \hy &=&  \frac {i (a_R \ba_I - a_{0I} \ba_R + 
\ba_R \wedge \ba_I)}
{\sqrt{(a_{0R})^{2} +\ba_R \cdot \ba_R}}\,\,.
\end{eqnarray}
We have $\hz \cdot \hy > 1$ by virtue of the fact that for a complex unit 
vector the square of the norm of the real part exceeds unity by an amount equal to 
the square of the norm of the imaginary part.

Let us define a real angle $\epsilon$ and real unit vector $\hk _r$ through
\begin{eqnarray}
\hx \cdot \hz &=& \cos (\epsilon/2)\,,\nonumber\\
\hx \wedge \hz &=& \sin (\epsilon/2)\,\, \hk _r\,.
\end{eqnarray}
These expressions remain  invariant under the transformation  $\epsilon/2 \to - 
\epsilon/2,\; \hk _r \to - 
\hk _r$, and hence we restrict the range of $\epsilon$ to $0 \le \epsilon/2 \le 
\pi$. Similarly, let us define a real positive  rapidity parameter  $\beta$ and a 
real unit vector $\hk _b$ through 
\begin{eqnarray}
\hz \cdot \hy &=& \cosh (\beta/2)\,,\nonumber\\
\hz \wedge \hy &=& i \sinh (\beta/2) \,\, \hk _b\,.
\end{eqnarray}
With these 
definitions the polar decomposition (3.5) can  be written as \begin{eqnarray}
S(a_0, \ba) &=& a_0 -i \ba \cdot \bs \nonumber\\
            &=& (\cosh (\beta/2) + \sinh (\beta/2)\,\, \hk _b \cdot \bs)
                (\cos (\epsilon/2) - i \sin (\epsilon/2)\,\, \hk _r \cdot \bs)\,.
\end{eqnarray}
The hyperbolic factor is manifestly hermitian positive definite with eigenvalues 
$\exp ( \pm \beta/2)$, and corresponds to a boost along  the spatial direction $\hk 
_b$  with rapidity parameter  $\beta$. The trigonometric factor is manifestly unitary 
and corresponds to spatial rotation by angle $\epsilon$ about the direction $\hk 
_r$. The suffix $b$/$r$ attached to $\hk$ signifies boost/rotation. Comparing 
(5.6), (5.7) with (5.5) we deduce that
\begin{eqnarray}
 \cosh (\beta/2) &=& \sqrt{(a_{0R})^{2} +\ba_R \cdot \ba_R}\,\,, \nonumber\\
  \tan (\epsilon/2) &=& \frac {\sqrt{\ba_R \cdot \ba_R}}{a_{0R}}\,\,,
\end{eqnarray}
and that the spatial rotation is about the direction of $(\mbox{sign}\,\,a_{0R})\ba 
_R$, while the boost is along the spatial direction 
 $(a_R \ba_I - a_{0I} \ba_R + \ba_R \wedge \ba_I)$.

In concluding this Section we wish to make the following remark. Since the set of 
all $2 \times 2 $ hermitian positive definite $SL(2,C)$ matrices $\exp ( 
\bx.\bs),\; \bx \in {\cal R}^3$ has the structure of ${\cal R}^3$ as a manifold, 
and since $SU(2)$ has the structure of $S^3$, it follows from the uniqueness of 
the polar decomposition that the $SL(2,C)$ group manifold has the structure of ${\cal 
R}^3 \times S^3$. In the case of $SO(3,1)\,[SO(3,C)]$, the role of $S^3$ will be  
played by the real projective space $RP^3 = S^3/Z_2$, consistent with the relation 
$SO(3,1) \sim SO(3,C) = SL(2,C) /Z_2$.

\section{Composition of Boosts: Wigner or Thomas Rotation}

As our second and final illustration of the theory of turns we have developed for 
the Lorentz group, we  apply it to the well known  
 problem of composition of Lorentz boosts. Let the first boost be in the spatial 
direction of 
the real unit vecor $\hm$, with rapidity parameter  $\beta _m$,  and let the 
second be in the 
direction of $\hn$, with rapidity parameter  $\beta _n$. The problem of 
composition of boosts  is trivial when 
$\hm$ and $\hn$ are parallel or antiparallel to one another, and so we assume 
$\hm$ and $\hn$ to be linearly independent. Let $(\hx_m , \hy_m )$ 
and $(\hx_n , \hy_n )$ be the turns corresponding to the two boosts. Then 
\begin{eqnarray}
\hx_m \cdot \hy_m &=& \cosh (\beta_m /2)\,, \nonumber\\ 
\hx_m \wedge \hy_m &=& i\sinh (\beta_m /2)\, \hm\,\,; \nonumber\\
\hx_n \cdot \hy_n &=& \cosh (\beta_n /2)\,,\nonumber\\ 
\hx_n \wedge \hy_n &=& i\sinh (\beta_n /2)\,\hn\,\,.
\end{eqnarray}
Let $\hz$ be the point where the head of the first turn and the tail of the 
second turn  meet (it turns out that $\hz$ is necessarily real). Since the pair $(\hx 
_m, \hy _m)$ is orthogonal to $\hm$ while 
the pair $(\hx _n, \hy _n)$ is orthogonal to $\hn$, the common meeting point $\hz$ 
has to be necessarily orthogonal to both $\hm$ and $\hn$:
  \begin{eqnarray} \hz = \pm 
\frac{\hm\wedge \hn} {\sqrt{(\hm\wedge \hn)\cdot(\hm\wedge \hn)}}\,\,.
\end{eqnarray}
Let $\hx$ be the tail of the first turn and $\hy$ be the head of the second turn 
when they so meet at $\hz$, so that the 
first boost can be represented by $(\hx,\hz)$ and the second by  
$(\hz,\hy)$. That is, $S(\hx _m, \hy _m) = S(\hx, \hz)$ and $S(\hx _n, \hy _n) = 
S(\hz, \hy)$. Solving for $\hx, \hy$ we have,
 \begin{eqnarray}
\hx&=& \cosh (\beta_m /2) \,\hz + i\sinh (\beta_m /2)\, \hz\wedge 
\hm\,\,, \nonumber\\
\hy&=& \cosh (\beta_n /2)\, \hz + i\sinh (\beta_n /2)\, \hn\wedge 
\hz\,\,.
\end{eqnarray}
We conclude from the parallelogram law  (4.5)  that  the product of the 
$\hm$ boost followed by the $\hn$ boost is 
\begin{eqnarray}
S(\hx_n, \hy_n) S(\hx_m, \hy_m) = 
S(\hz, \hy) S(\hx, \hz)
=S(\hx, \hy)\,.
\end{eqnarray}

To gain insight into the product $S(\hx , \hy) \in SL(2,C)$, let us write it in the 
form  $S(\hx, \hy) = a_0 -i\ba.\bs$, where 
\begin{eqnarray}
a_0 = \hx\cdot\hy & \equiv & a_{0R} + i a_{0I}\,, \nonumber\\
\ba = \hx\wedge \hy &\equiv & \ba _R + i \ba _I\,.
\end{eqnarray}
Use of (6.3) yields the following expressions for the real and imaginary parts of 
$a_0, \ba $: 
\begin{eqnarray}
a_{0R} &=&  \cosh (\beta_m /2)\cosh (\beta_n /2)\nonumber\\
&&\;\;\;+  \sinh (\beta_m /2) \sinh (\beta_n /2)\,\,\hm\cdot \hn\,,\nonumber\\
 a_{0I} &=& 0\,; \nonumber\\
\ba_R &=& \sinh (\beta_m /2) \sinh (\beta_n /2) 
\hm\wedge\hn\,,\nonumber\\
\ba_I &=&\sinh (\beta_m /2) \cosh (\beta_n /2)\,\, \hm
+ \sinh (\beta_n /2)\cosh (\beta_m /2)\,\,\hn\,.\nonumber\\
\end{eqnarray}
The vector parameter $\ba$ in (6.6) determining the product of the  $\hm$ and 
$\hn$ boosts is  complex, as was to be expected,  since the product of two 
Lorentz boosts in neither a (pure) boost nor a spatial rotation. 

We may
 now apply the  polar decomposition developed in the last Section to the product  
of the two boosts: 
\begin{eqnarray}
S(\hx_n, \hy_n) S(\hx_m, \hy_m)& =& 
 a_0 - i \ba \cdot \bs         \nonumber \\
& =& S(\hz^\prime, \hy^\prime)~S(\hx^\prime , \hz^\prime)\,, 
\end{eqnarray}
where we require  $S(\hz^\prime, \hy^\prime)$  to correspond to a boost and 
$S(\hx^\prime , \hz^\prime)$ to a spatial rotation, and  $a_0,\ba$ are given by 
(6.6). The spatial 
rotation so obtained  is known as the {\it Wigner} or {\it Thomas 
Rotation}\cite{thomas}. 

It follows from (5.5) and (6.6)  that the Wigner rotation $S(\hx^\prime , 
\hz^\prime)$  is about the direction $\ba_R \sim 
\hm \wedge \hn$. That is, the Wigner rotation is in the plane spanned by $\hm$ and 
$\hn$.  Let $\epsilon$ be the magnitude of the  Wigner rotation.
 Then from (5.9) we have
\begin{eqnarray}
\tan(\epsilon/2) &=& \sqrt{(a_{0R})^{2} +\ba_R \cdot \ba_R} \nonumber\\
                &=& \frac{\sin\theta}{\kappa +\cos\theta}\,\,, 
\end{eqnarray} 
where $\kappa=\coth (\beta_n /2)\coth (\beta_m /2)$, and $\theta$ is the angle 
between $\hm$ and $\hn$.

Having computed the Wigner rotation part, we now examine the boost part 
 $S(\hz^\prime, \hy^\prime)$. 
 Let $\beta_{{\rm res}}$ denote the rapidity  parameter for this boost. Then one 
finds from  (5.9) 
\begin{eqnarray}
\cosh\beta_{{\rm res}} &=&  2\left((a_{0R})^{2} +\ba_R \cdot \ba_R \right) -1 
\nonumber\\ &=&  
\cosh \beta_m\cosh \beta_n + \sinh \beta_m\sinh \beta_n\cos\theta\,.
\end{eqnarray}
Further, it follows from (5.5) and (6.6) that this resultant boost is in the 
direction of
 \begin{eqnarray}
     -i\hz^\prime \wedge \hy^\prime  &=&  \frac { (a_R \ba_I - a_{0I} \ba_R + 
        \ba_R \wedge \ba_I)}{\sqrt{(a_{0R})^{2} +\ba_R \cdot \ba_R}}\nonumber\\
              &=&\frac{1}{2} \cosh \beta_m\sinh\beta_n\,\, \hn \nonumber\\
                &&\;\;+\frac{1}{2}\sinh \beta_m\,\,\hm 
                + \sinh^2 (\beta_n/2)\sinh \beta_m\cos\theta\,\,\hn\,.\nonumber\\ 
\end{eqnarray}
Thus  the  resultant  boost is about a direction  in the plane spanned by $\hm$ and 
$\hn$. If $\phi$ is the angle between the resultant boost and $\hn$,   one 
finds that 
\begin{eqnarray}
\tan\phi = \frac{\sin\theta \sinh\beta_m }{\cosh\beta_m \sinh\beta_n +
\cos\theta \cosh\beta_n \sinh \beta_m}\,\,.
\end{eqnarray}
Thus the product of two Lorentz boosts of rapidity parameters  $\beta_m, \beta_n$ 
 and spatial directions $\hm, \hn$ is a Wigner rotation of 
amount $\epsilon$ in the plane spanned by $\hm, \hn$ followed by a Lorentz boost 
of rapidity parameter  $\beta_{{\rm res}}$  along the direction that lies in the 
plane spanned 
by $\hm, \hn$ and makes  an angle $\phi$ with $\hn$. The expressions for $\epsilon, 
\beta_{{\rm res}}$ and $\phi$ derived using turns are consistent with known 
results\cite{results}, but our aim of this exercise was simply to illustrate the 
theory of turns we have developed for the Lorentz group.

\section{Concluding Remarks}
We have examined in detail the ideas underlying Hamilton's theory of turns 
for $SU(2)$, in a manner that paves the way to developing a similar geometrical 
construction for other groups. After a brief review of the theories of turns for 
$SU(2)$ and $SU(1,1)$, we  worked out a theory of turns for the group 
$SL(2,C)$. Formulae for the polar decomposition of an $SL(2,C)$ element were  
derived within the framework of the theory of turns  developed here, and were  put to 
use to compose Lorentz boosts and to calculate the resulting Wigner rotation. 
These exercises, carried out for illustration of the geometric construction, led 
to acceptable results. Decomposition of an arbitrary element of the Lorentz group in 
the form spatial 
rotation--followed by boost in a fixed direction--followed by spatial rotation 
  can be analysed in a similar manner using turns. 

It is gratifying to note that, 
with our earlier 
generalization\cite{prl,jmp} of the theory of turns to $SU(1,1) \sim  SL(2,R) 
=Sp(2,R)$,  
the present  generalization  renders a geometric representation in the spirit of 
Hamilton  available to all low dimensional simple Lie groups of interest in physics.

The geometry of turns for $SU(2)$ is applicable to any problem that involves the 
group $SU(2)$. Two-level systems constitute an important class of such problems, 
but not the only ones. As remarked earlier, the theory of turns has led to the 
formulation and solution of important sythesis problems in these 
contexts\cite{pramana,gadget1,gadget2}. 
For instance, we have  the result that  {\em all}  linear intensity preserving 
transformations of polarization optics can be synthesised using just two 
quarterwave plates and one halfwave plate\cite{gadget2,gori}. Notwithstanding its 
wording, this 
result is applicable not only to polarization optics, but also to other systems 
involving $SU(2)$. For instance, in the case of nuclear magnetic resonance $\pi/2$ 
and $\pi$-pulses will correspond  respectively to the same $SU(2)$ transformations 
 as quarter and halfwave  plates  in polarization optics. (These are respectively 
the eighth and fourth symmetric roots of  the unit matrix.)  And hence 
the above result in this context  will read: any unitary evolution ($\;SU(2)$ 
transformation )  of 
a spin half system can be achieved using two $\pi/2$ pulses and one $\pi$ pulse. 
 Further, the theory of $SU(2)$ turns has helped to clarify the deep relationship 
between the structure of this group and geometric phase in two-level 
systems\cite{jphys}. It 
should be of interest to carry out similar geometric studies for the 
Lorentz group  using the theory developed here.

In this work we have restricted ourselves to simply demonstrating that a consistent 
theory of turns is possible for the Lorentz group. The applications considered 
served the limited purpose of showing the effectiveness of the theory. More 
elaboration will, of course, be needed to handle the questions raised in the last 
paragraph, and we plan to return to these and other issues elsewhere. \\

{\bf Acknowledgement: } The authors wish to thank  Dr. S. Arun Kumar for 
his assistance  in preparing the illustration.


\begin{references}

\bibitem{hamilton} 
W. R. Hamilton, {\em Lectures on Quaternions} (Dublin, 1853).

\bibitem{biedenharn}
L. C. Biedenharn and J. D. Louck, {\em Angular momentum in quantum physics}, in 
 \underline{Encyclopedia of Mathematics} \underline{and its Applications} 
(Addison-Wesley, Reading, MA, 1981).

\bibitem{prl}
R. Simon, N. Mukunda, and E. C. G. Sudarshan, Phys. Rev. Lett. 
{\bf 62}, 1331(1989).

\bibitem{jmp}
R. Simon, N. Mukunda, and E. C. G. Sudarshan, J. Math. 
Phys. {\bf 30}, 1000 (1989).   

\bibitem{pramana}
 R. Simon, N. Mukunda, and E. C. G. Sudarshan, 
Pramana -- J. Phys. {\bf 32}, 769 (1989).  

\bibitem{gadget1}
 R. Simon and N. Mukunda,  Phys.  Lett.  A {\bf 138, } 474 (1989).


\bibitem{gadget2} 
 R. Simon and N. Mukunda, Phys.  Lett. A  {\bf 143, } 165 (1990). 


\bibitem{gori}
V. Bagini, R. Borghi, F. Gori, M. Santarsiero, F. Frezza, G. Schettini, and 
 G. S. Spagnolo, Eur. J. Phys. {\bf 17}, 279 (1996).


\bibitem{geophase}
M. V. Berry, Proc. Roy. Soc. London A {\bf 392}, 45 (1984); 
S.Pancharatnam, Proc. Ind. Acad. Sci. A {\bf 44}, 247 (1956).



\bibitem{jphys}
 R. Simon, N. Mukunda,  J.  Phys.  A: Math.  Gen.  {\bf 25, } 6135 (1992).


\bibitem{orbits}
N. Mukunda, R. Simon and E. C. G. Sudarshan,  Ind.  J.  Pure and Appl.  
Math.  {\bf 19, } 91 (1988). 


\bibitem{lorentz}
See, for example, I. M. Gelfand, R. A. Minlos, and Z. Ya. Shapiro, 
{\em Representations of the rotation and the Lorentz groups and their 
applications} (Pergamon, New York, 1963). Further references can be found in, 
 N. Mukunda and R. Simon,  J.  Math.  Phys.  {\bf 36, } 5170 (1995).

\bibitem{thomas}
L. H. Thomas, Nature {\bf 117}, 514 (1926); Phil. Mag. {\bf 3}, 1 (1927);  
E. P.  Wigner, Ann. Math. {\bf 40}, 149 (1939). An illuminating discussion can be 
found in the text book of H. Goldstein, {\em Classical Mechanics}, 2nd Edn. 
(Addison-Wesley, Reading, MA 1980),  p.286.




\bibitem{results}
A. A. Ungar, Found. Phys. Lett. {\bf 1}, 57 (1988), where further references can 
be found. 
 R. Simon and N. Mukunda,  Found.  Phys.  Lett.  {\bf 3, } 425 (1990);
N. Salingaros, J. Math. Phys. {\bf 27}, 157 (1986);
Y.S. Kim and D. Son,  J. Math. Phys. {\bf 27}, 2228 (1986);
W. E. Baylis and G. Jones, J. Math. Phys. {\bf 29}, 57 (1988);
P. K. Aravind,  Am. J. Phys. {\bf 65}, 634 ( 1997), and references 
therein.              




 \end{references}
\end{document}